\documentstyle[prb,twocolumn,aps,epsf]{revtex}
\begin{document}

\twocolumn[
\hsize\textwidth\columnwidth\hsize\csname@twocolumnfalse\endcsname

\title{Plasmon dispersion in dilute 2D electron systems: 
Quantum-Classical and Wigner Crystal-Electron Liquid Crossover}
\author{E. H.\ Hwang and S. Das Sarma}
\address{Department of Physics, University of Maryland, College Park,
Maryland  20742-4111 } 
\date{\today}
\maketitle

\begin{abstract}

We theoretically calculate the finite wave vector plasmon dispersion in 
a low density 2D electron layer taking 
into account finite temperature,  finite layer width, and 
local field corrections.
We compare our theoretical results with recent Raman scattering 
spectroscopic experimental 2D plasmon dispersion data 
in GaAs quantum wells at very low 
carrier densities ($r_s > 10$) and large wave vectors ($q \ge k_F$). 
We find good agreement with the experimental data, providing an 
explanation for why the experimentally measured dispersion seems to 
obey the simple classical long wavelength 2D plasmon dispersion formula.
We also provide a critical discussion on the observable
manifestations of the quantum-classical {\it and} the Wigner 
crystal - electron
liquid crossover behavior in the 2D plasmon properties as a function of
electron density and temperature in GaAs quantum well systems.


\end{abstract}
\vspace{0.2in}
]

\newpage

A plasmon (or, a plasma excitation) is a fundamental elementary 
excitation of an electron system (or, for that matter, any charged 
particle system). It is the collective (normal) mode of charge density 
oscillation in the free carrier system, which is present both in 
classical and quantum plasmas. Studying the collective plasmon 
excitation in the electron gas has been among the very first theoretical 
quantum mechanical many body problems studied in solid state physics 
dating back to the early 1950s \cite{one,two,three}. 
A great deal of theoretical and 
experimental work has been carried out during the last fifty years on 
the issue of observable many body effects in the plasmon dispersion 
relation, originally in the context of bulk metals \cite{four}
and more recently for two dimensional (2D) plasmons \cite{five}
occurring in 2D electron gases confined 
in semiconductor inversion layers, heterojunctions, quantum wells, and 
superlattices.  The specific theoretical issue is that the plasmon 
frequency ($\omega_p$) is exactly known only at long wavelengths 
($q \rightarrow 0$) where the $f$-sum rule fixes the plasma frequency 
to be necessarily that given by classical electrodynamics, namely, 
$\omega_p = (2\pi n e^2/m)^{1/2} \sqrt{q}$ in 2D and 
$\omega_p=(4\pi n e^2/m)^{1/2}$ in 3D where $n$ is the 2D (3D) electron 
density and $e^2$ is interpreted as $e^2 \rightarrow e^2/\kappa$ where 
$\kappa$ is the background lattice dielectric constant. At finite $q$, 
away from the long wavelength limit, there are several corrections to 
the plasmon dispersion $\omega_p(q)$ arising from nonlocal 
(finite wave vector) response, finite temperature 
thermal corrections, many-body effects, 
local field corrections, and other mechanisms 
relevant to the specific electron system being 
studied (e.g., finite width of the 2D layer, band structure effects, 
interface effects, mode coupling effects to phonon and other possible 
elementary excitations, impurity scattering or disorder effects, etc.). 
Significant deviations from the long wavelength classical result are 
predicted to occur \cite{five,six,seven,eight,nine,ten}
in 2D (as well as in 3D \cite{four} --- in this paper we restrict 
our considerations, motivated by recent experiments\cite{eleven}, 
entirely to 2D 
plasmons) plasmon dispersion as the electron density is lowered, and 
the plasmon wave vector ceases to be in the long wavelength limit 
($q \ll k_F$, where $k_F=(4 \pi n/g)^{1/2}$, with $g$ as the degeneracy 
factor of the 2D electron system including spin, is the 2D Fermi wave 
vector) and starts becoming comparable ($q \approx k_F$) to the Fermi 
wave vector. Using the usual dimensionless electron gas (quantum) coupling 
strength parameter $r_s$ ($\equiv 1/a_B \sqrt{\pi n}$, where 
$a_B$ ($\equiv \hbar^2/me^2$) is the effective Bohr radius), 
where $r_s$ is the ratio of the noninteracting kinetic energy 
to the average potential energy at $T=0$,
one expects large finite wave 
vector corrections to the classical long wavelength plasmon dispersion 
relation $\omega_{cl}(q) = \omega_0 \sqrt{q}$, with 
$\omega_0^2 = 2 \pi ne^2/m$, for $r_s \gg 1$ where 
the electron gas is strongly interacting. In fact, there are theoretical 
predictions, based on model many-body plasmon dispersion 
calculations \cite{ten}, that the 2D plasmon frequency may 
renormalize to zero at finite wave vectors (and $r_s \gg 1$) 
due to interaction-induced softening of the plasmon dispersion. 
In general, theories including higher order many-body and other 
effects predict a lowering of the plasma frequency below the classical 
plasmon dispersion curve for $r_s > 1$ and finite values of $q/k_F$.

It therefore comes as a surprise that recent inelastic light (Raman) 
scattering based direct measurements \cite{eleven} of 2D plasmon 
dispersion in high quality low density (down to electron densities as 
low as $n \approx 5 \times 10^8 cm^{-2}$ corresponding to $r_s$ parameter 
as high as 25) electron systems in GaAs quantum wells  find 
remarkably good agreement between the measured 2D plasmon dispersion 
and the simple classical formula ($\omega_p \approx\omega_{cl}$) upto wave 
vectors as large as $q \sim 2k_F$. Earlier experimental measurements 
\cite{twelve} of 2D plasmon dispersion in semiconductor structures were 
typically restricted to higher carrier densities ($r_s \le 1$) and lower 
wave vectors ($q \sim 0.1-0.2k_F$) where the classical $q^{1/2}$ plasmon 
dispersion relation is valid by virtue of the applicability of the long 
wavelength $f$-sum rule. A primary motivation for the recent low electron 
density 2D plasmon dispersion measurement \cite{eleven}, which we are 
trying to theoretically understand in this paper, has been to explore 
high-wave vector ($q \ge k_F$) and low density ($r_s \gg 1$) dispersion 
corrections to the classical $q^{1/2}$ 2D plasmon dispersion formula. 
The unexpected finding \cite{eleven} that the experimental 2D plasmon 
dispersion follows quantitatively the classical $q^{1/2}$ formula upto 
large values of $q/k_F$ ($\sim 2$) even for very strongly interacting 
($r_s \sim 25$) 2D electron systems is a significant puzzle, particularly 
in view of the extensive existing theoretical 
literature \cite{five,six,seven,eight,nine,ten} on finite wave 
vector many-body and nonlocal plasmon dispersion corrections showing very 
large deviations from the classical plasma frequency.

By carrying out a realistic random phase approximation (RPA) calculation 
of the finite temperature, finite wave vector 2D plasmon dispersion in 
the actual GaAs quantum wells used in ref. \onlinecite{eleven}, we 
provide a partial resolution of the puzzle posed by the data presented 
in ref. \onlinecite{eleven}. In particular, we show that RPA provides an 
excellent quantitative description of the data --- the nonlocal finite 
$q$ corrections introduced by RPA (which are substantial for 
$q/k_F \sim 2$) are not explicitly manifest in the experimentally measured 
plasmon dispersion \cite{eleven} because of a fortuitous
cancellation among a number of independent contributions to the plasmon 
energy, mostly the almost exact accidental numerical calculation between 
the higher-order (i.e., $q^{3/2}$ and higher) nonlocal finite wave vector 
correction (tending to increase the plasma frequency above the classical 
dispersion curve and becoming quantitatively prominent for $q/k_F > 1$) 
and the finite layer width correction arising from the {\it quasi}-2D 
nature of the electron layer (which by itself lowers the plasma 
frequency below the classical dispersion curve because the usual 
$q^{-1}$ Coulomb interaction in a strictly 2D system is softened 
in the {\it quasi}-2D system, and this softening induced lowering of the 
plasma frequency is particularly quantitatively significant in low 
density systems, where the layer width in the transverse direction is 
large due to the weakening of the Hartree selfconsistent potential 
contribution to the confinement). 
We also find, as explained below, that a slight ($\sim 10 \%$) adjustment 
in the quoted sample densities in ref. \onlinecite{eleven}, 
where density has not been independently measured but inferred from 
the measured plasmon dispersion, significantly improves the agreement 
between our theoretical results and experimental data.

The plasmon dispersion can be calculated by finding poles 
of the density-density correlation function.
Within RPA (or its simple generalizations \cite{seven,nine,ten} 
including local field corrections arising from correlation effects) 
plasmon modes at finite wave vectors and finite temperatures
are given by the zeros of the dielectric function,
$\epsilon(q,\omega;T) = 1 - v(q)\Pi_0(q,\omega;T)$, where $v(q)$ is the
Coulomb interaction modified by both the quasi-2D form factor due to the
finite width of the 2D quantum well and correlation-induced 
local field effects, and $\Pi_0(q,\omega,T)$ is the 
noninteracting 2D finite temperature irreducible polarizability.
The modification of the Coulomb interaction due to correlation
induced local field correction is modeled by a static correlation 
factor $G(q)$, which we calculate within the Hubbard approximation
\cite{four,seven,ten}. In the presence of local field corrections 
the Coulomb interaction is modified in the following manner:
$v(q)\rightarrow v(q)[1-G(q)]$. Note that local field corrections,
$0<G(q)<1$, tend to soften the Coulomb interaction because 
exchange-correlation effects tend to keep the electrons away from 
each other reducing the effective Coulomb interaction.
We calculate numerically the plasmon dispersion by solving 
$\epsilon(q,\omega;T)=0$
to obtain $\omega_p(q;n,T)$ by incorporating
thermal, finite thickness, and local field  correlation effects. 
(All other effects, e.g. phonons, give
negligible  corrections to the plasmon dispersion in a low
density GaAs quantum well electron system.)
For each density the chemical potential of the 2D system has to be 
calculated self-consistently at temperature $T$ since at the low 
densities of interest to us, where $E_F \sim k_BT$, the chemical potential 
is very different from the Fermi energy.
In our calculation we incorporate finite thickness corrections by using 
a quasi-2D form factor appropriate for 
an infinite square well potential with a width $d$. At the low 
electron densities used in ref. \onlinecite{eleven} the infinite 
square well confinement model is a very good approximation
since the Hartree selfconsistent effects are weak.
We include the local-field correction \cite{seven,ten} 
using the Hubbard approximation (HA), where the 2D HA correlation factor is 
given by 
\[G(q)=\frac{1}{g}\frac{q}{\sqrt{q^2+k_0^2}},\]
where $k_0=k_F$ at $T=0$.
We generalize the $T=0$ HA \cite{seven,ten} to finite temperatures by
simply redefining $k_0$ to be the finite temperature analogy of the 
Fermi wave vector using the formula:
$k_0(T)=k_F (T/T_F) \ln (1+ e^{\mu/k_BT})$, where 
$\mu=\mu(T)=k_BT\ln[e^{(n/k_BT)(2\pi/gm)}-1]$
is the finite temperature 2D chemical potential obtained
from the total 
number of particles $n=\int dE D(E) f(E)$, using the 2D electron density
of states, $D(E)$, and the Fermi distribution function, $f(E)$.
The use of this particular (i.e., HA) local field correction is an
uncontrolled approximation of our theory since there is no generally 
accepted theoretical framework to incorporate correlation effects 
(beyond RPA) in the plasmon dispersion calculation. Our finite 
temperature generation of the HA is, however, quite reasonable 
since it correctly interpolates between the $T=0$ 2D HA \cite{seven} 
with $k_0(T=0)=k_F$ and the high temperature result, $k_0(T/T_F 
\rightarrow \infty)=0$, of vanishing local field correction.
We have checked that the use of other static local field corrections (e.g, 
the finite-T STLS corrections \cite{r12}) do not significantly alter our
conclusion. It is well-known \cite{four,seven} that the HA for
the local field correction, while being static and theoretically 
uncontrolled, has the great advantage of being simple and reasonably 
quantitatively accurate compared with other local field correction 
approximations.

In Fig. 1 we show our calculated plasmon dispersion along with 
the available experimental data from ref. \onlinecite{eleven}. 
In Figs. 1(a) and 1(b) respectively plasmon dispersion curves for $r_s=5.2$ 
($n=1.2 \times 10^{10} cm^{-2}$) and  
$r_s = 17.2$ ($n=1.1 \times 10^{9} cm^{-2}$) are shown at $T=200 mK$.
We use in our calculations system parameters corresponding to 
GaAs-Al$_x$Ga$_{1-x}$As quantum wells with a well width $d=330\AA$ 
as appropriate for ref. \onlinecite{eleven}.
Thin solid lines represent the classical long wavelength 
plasmon dispersion ($\omega_{cl}=\omega_0 \sqrt q$). Dotted
lines show the dispersion including finite wave vector
non-local (i.e., higher order in $q$)
effects calculated within RPA, which leads
to an increase in plasma frequency compared with $\omega_{cl}$.
The leading order RPA dispersion correction is given by 
$\omega_p(q)/\omega_{cl} =
1+(3/4)(q/q_{TF})$, where $q_{TF}$ ($=gme^2/\hbar^2$) 
is the 2D screening wave vector. Dashed lines are the 
dispersions including corrections by both non-local and finite thickness
effects.
Finite well width reduces the plasma frequency by softening the
2D Coulomb interaction, and its long
wavelength dispersion in an infinite square well potential model 
can be calculated by straightforward algebra to be  given by
$\omega_p(q)/\omega_{cl} = 1- 0.207(d/a_B)(q/q_{TF})$. 
Noting that $g=2$ (spin degeneracy) in GaAs quantum wells, i.e. 
$q_{TF}a_B=2$, we find that the two dispersion corrections (non-local and 
finite width) approximately cancel each other when $3q/4q_{TF}=0.207 
dq/a_Bq_{TF}$. Thus we find that
non-local effects and finite thickness corrections cancel each other 
almost exactly in a
quantum well with well thickness $d\approx 3.6 a_B$. The GaAs 
($a_B\approx 90 \AA$) sample used in
ref. \onlinecite{eleven} has a quantum well width $d=330\AA$, which 
gives rise to an almost exact cancellation between non-local effects 
and finite thickness corrections. 
As shown in Fig. 1 RPA results with realistic finite
thickness effects agree with the classical formula upto a large 
wave vector ($q\sim 2k_F$) and 
provide reasonable  quantitative agreement with 
experimental 2D plasmon dispersion data. 
Dot-dashed lines in Fig. 1 are the plasmon dispersion curves
including local field corrections, 
non-local dispersion, and finite thickness effects at $T=0$. 
The local field corrections tend to reduce the plasma frequency
since it softens the Coulomb interaction by keeping the 
electrons effectively away from each other, and
in the long wavelength limit the 2D plasmon dispersion within the 
HA local field corrections is given at $T=0$ by 
$\omega_p(q)/\omega_{cl} = 1 - (r_s/2\sqrt{2})(q/q_{TF})$. As the density
decreases (or $r_s$ increases) the reduction of the plasma frequency
due to local field corrections is stronger, and this effect 
dominates in large wave vectors.
Thick solid lines show the full plasmon dispersion including finite
temperature effects at $T=200mK$ including all the corrections described
above. (Long dashed line represents the plasmon dispersion including
all corrections at $T=200mK$ with zero temperature HA, i.e., $k_0=k_F$.) 
Even within RPA 
finite temperature effect by itself increases
the plasma frequency. The thermal
enhancement is negligible (exponentially small for $T/T_F \ll 1$) at low $T$
in high density GaAs 2D electron systems since
the Fermi temperature in these systems is much greater than the experimental 
temperature ($T/T_F \ll 1$). However,  a very low density
electron system (e.g., 
electrons on the surface of liquid Helium, where the Fermi temperature is 

\begin{figure}
\epsfysize=2.4in
\centerline{\epsffile{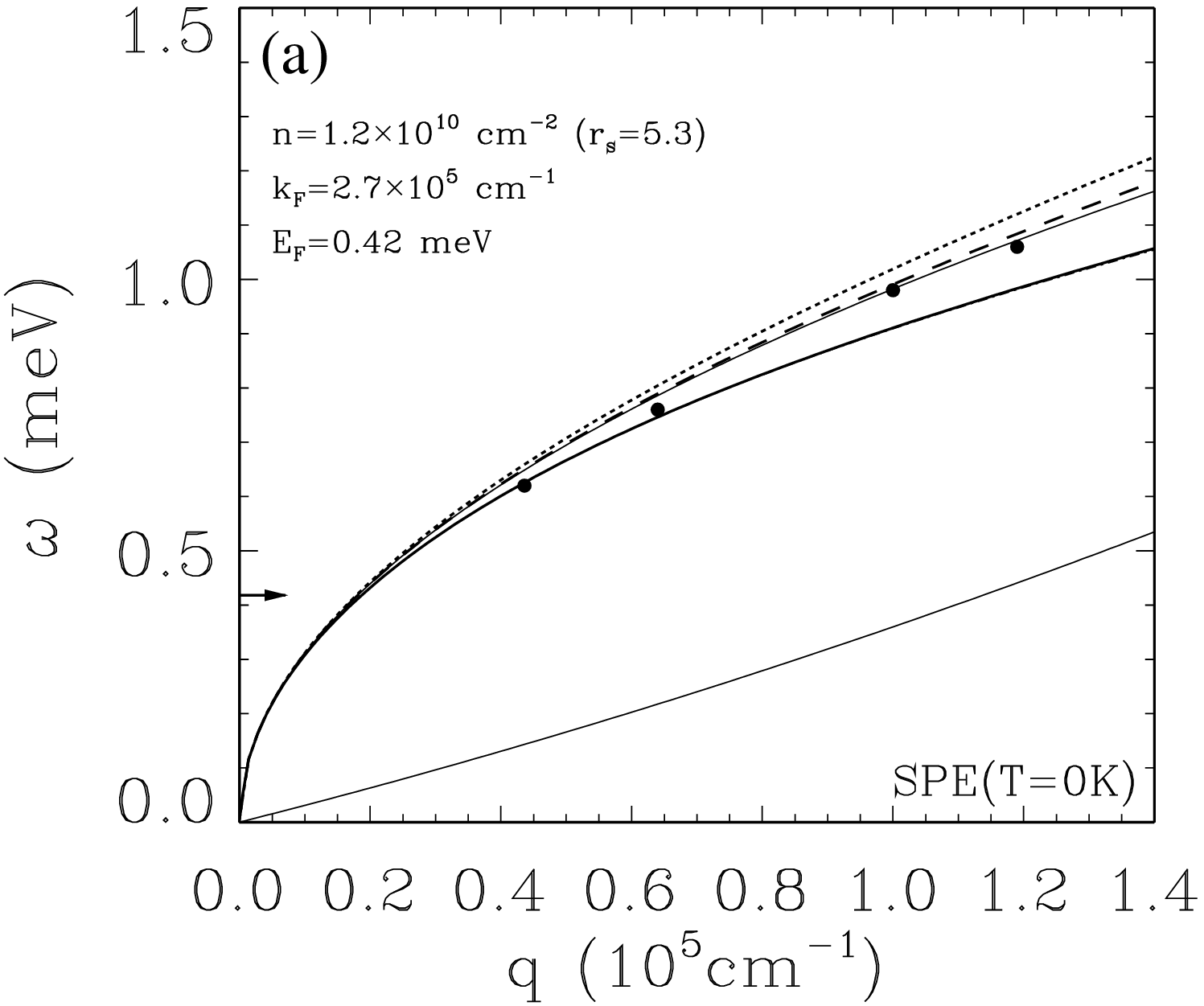} }
\vspace{0.5cm}
\epsfysize=2.4in
\centerline{\epsffile{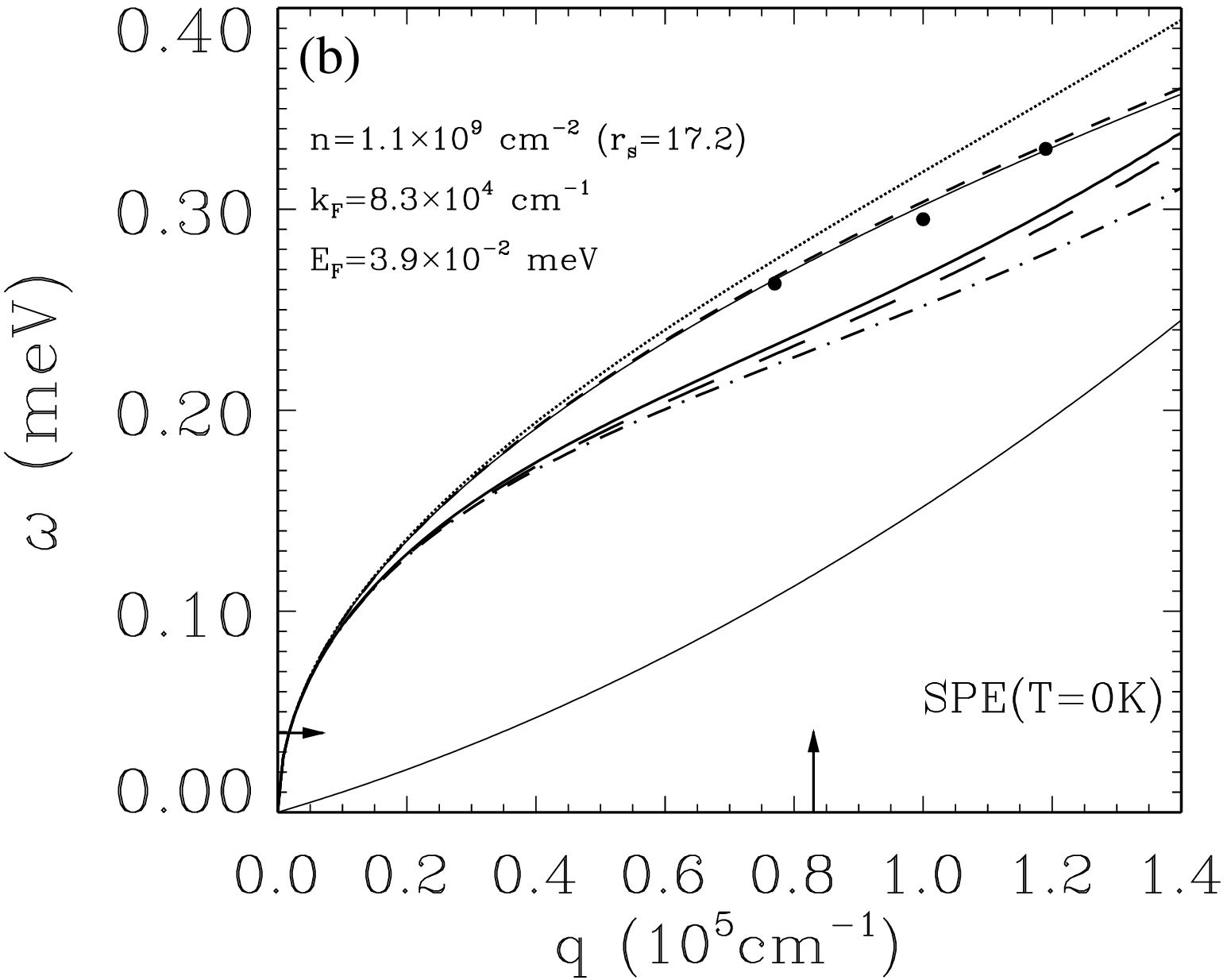}}
\vspace{0.5cm}
\caption{
Calculated plasmon dispersions with available 
experimental data [11] for (a) $n=1.2 \times 10^{10} cm^{-2}$ ($r_s=5.3$)
and (b) $n=1.1 \times 10^{9} cm^{-2}$ ($r_s = 17.2$).
Thin solid line is 
the classical local plasmon dispersion
and thick solid line indicates the dispersions with all corrections 
(as explained in the text) at a finite temperature ($T=200 mK$). 
Other lines in the figure are described in the text.
The arrows indicate the Fermi wave vector and Fermi energy.
The dots are experimental data points from ref. [11].}
\label{fig1}
\end{figure}

\noindent
smaller than or comparable with experimental temperature)  
is essentially a non-degenerate classical plasma,
where the long wavelength plasmon dispersion is given by 
$\omega_p(q)/\omega_{cl} = 1 + (3/2\sqrt{2})(T/T_F)(q/r_s)$.
In a quantum 
plasma ($T\ll T_F$) on the other hand the leading order 
thermal correction to the $T=0$ plasmon dispersion is exponentially 
weak in temperature. We thus see that for $T/T_F$ not too small 
(which is the experimental situation in ref. \onlinecite{eleven}, 
where $T/T_F \approx 0.4$ at the lowest density) the thermal correction 
to the plasma frequency tends to cancel the local field correction 
similar to the cancellation between the nonlocal finite $q$ corrections 
and quasi-2D finite width corrections.

In Fig. 1(a) the plasmon dispersion for $r_s=5.3$ 
($n=1.2\times 10^{10} cm^{-2}$) is shown. 
In this relatively high density system the experimental
data lie in the long wavelength 
($q < k_F = 2.7 \times 10^5 cm ^{-1}$) and the low temperature 
($T\ll T_F = 5.0K$) limit. 
Note that the enhancement of plasma frequency by non-local effects 
is almost cancelled by finite thickness effects.
The local field corrections and finite
temperature effects are not important in this sample since these 
effects are quantitatively significant only at low $n$ or large $r_s$.
In Fig. 1(b) we show the plasmon dispersion for $r_s = 17.2$
($n=1.1 \times 10^{9} cm^{-2}$). Experiment data \cite{eleven} for 
this sample lie in the effective
large wave vector ($q > k_F = 0.83 \times 10^5 cm ^{-1}$) regime. 
Even in this low density sample 
(and at large effective wave vectors) the experimental data 
can be seen to agree very 
well with the classical plasma dispersion and the finite 
thickness RPA calculations. Since the non-local effects are 
almost canceled by the finite width correction, we
speculate that the reduction of the plasma frequency due to local
field corrections is perhaps almost exactly
canceled by the thermal enhancement of the plasma frequency.
Our calculated results (Fig. 1(b)), however, 
show the local field corrections to be  too large 
in this low density sample to cancel
out with thermal effects. 
It is certainly possible, perhaps even likely, that our calculated 
HA local field corrections
are strong overestimations of the actual finite temperature 
($T/T_F \sim 0.4 $ in the experiment) local field corrections, and in reality
local field corrections, being much smaller that our HA results,
do in fact cancel out with the finite temperature 
enhancement. This remains an important open question for future theoretical 
work. One would need a quantitatively more accurate and
theoretically well-controlled finite temperature theory
for local field corrections for this purpose. At present no
such local field theory exists in the theoretical literature.

Since the local field corrections (even within HA)
are regarded as improvement to RPA in calculating 
many physical properties of low density systems (strongly correlated
systems) the large discrepancy between the plasmon 
mode dispersion (in Fig. 1(b)) including local field effects and the
experimental data in this low density electron system 
is unexpected.
It is troublesome to uncritically accept that correlation effects
are negligible based on the mere speculation that finite temperature 
local field corrections may be small (in other words, much
smaller than what we calculate within the HA).
We now suggest another possibility. 
In ref. \onlinecite{eleven} the electron densities are estimated
using the classical local plasmon dispersion formula 
based on the fact that the $\sqrt{q}$ dispersion seems to apply
very well to the experimental plasma dispersion. This could,
however, be problematic, and may in fact lead to an underestimation of $n$.
We now compare the zero temperature classical local
plasmon formula used to estimate the electron densities given 
in ref. \onlinecite{eleven} 
($n=1.1$ and $3.7 \times 10^{9} cm^{-2}$) with
the corresponding finite temperature plasmon dispersion including all 
the corrections described above, using somewhat higher densities in the 
compete dispersion calculation.
In Fig. 2 we show the plasmon dispersion including all corrections
for densities

\begin{figure}
\epsfysize=2.4in
\centerline{\epsffile{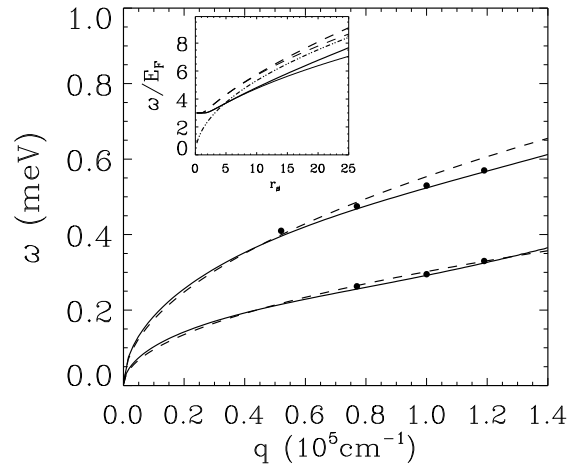} }
\vspace{0.5cm}
\caption{
Comparison of finite temperature ($T=200mK$) plasmon 
dispersions (solid lines) including all corrections 
for the density $n=1.37, 4.1 \times 10^{9} cm^{-2}$ 
with zero temperature
classical local dispersions (dashed lines) for 
$n=1.1, 3.7 \times 10^{9} cm^{-2}$.
Experimental data are taken from ref. [11].
Inset shows the plasma frequency as a function of density parameter
$r_s$ at $q=k_F$.
Dot dashed line is the classical plasmon mode, dashed lines are RPA
modes (thick line for $T=300mK$ and thin line for $T=0K$), and solid lines
are HA modes (thick line for $T=300mK$ and thin line for $T=0K$).
}
\label{fig2}
\end{figure}

\noindent
$n=1.37$ (lower solid line) and $4.1 \times 10^{9} cm^{-2}$ 
(upper solid line). Experimental data points 
are taken from ref. \onlinecite{eleven}. Dashed lines in Fig.2 are the
classical $T=0$ local plasmon dispersion for the density 
$n=1.1$ (lower line) 
and $3.7 \times 10^{9} cm^{-2}$ (upper line).
Our calculated full plasmon dispersion in Fig. 2
for somewhat higher densities including
all corrections also agrees very well with experiment just as the classical 
$T=0$ formula apparently does for the lower densities 
proposed in the experiment.
Thus, whether the local field corrections 
are large or not is not obviously clear until one can measure
the experimental electron density using an independent method.
This is, however, very difficult to do in the interesting regime of
very low electron densities where the usual Hall density measurement 
fails. At this stage all we can say is that the full 
plasmon dispersion (including local field corrections and the other
effects) could explain the experimental data provided one 
uses somewhat larger  electron densities than the experimental 
carrier density estimates (based on a comparison of the
data with the classical $T=0$ plasma dispersion formula) 
in ref. \onlinecite{eleven}.
Inset in Fig. 2 shows the quantum-classical crossover of the plasma
frequency at wave vector $q=k_F$ and $T=300 mK$ as a function of the 
density parameter $r_s$. In the high density limit the enhancement of 
the plasma frequency by non-local
effects dominates all other corrections, but in the low density limit  
the local field corrections give rise to a large decrease of 
the plasma frequency. Thermal correction of the plasmon mode 
also increases as the density decreases.

\begin{figure}
\epsfysize=2.4in
\centerline{\epsffile{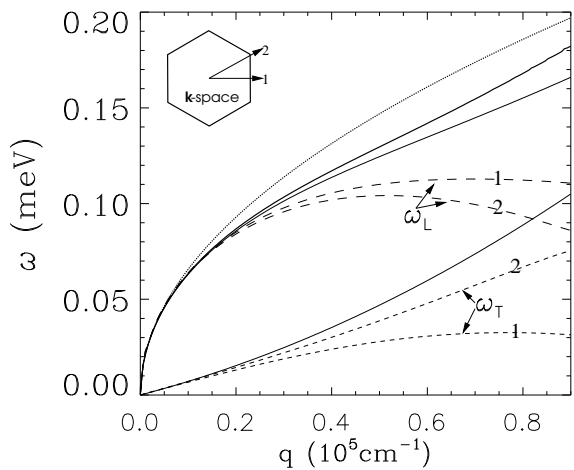} }
\vspace{0.5cm}
\caption{
Comparison of plasmon dispersions calculated within HA (solid lines, the thick
and thin line being for $T=200mK$ and $T=0K$, respectively)
with the phonon modes of a hexagonal Wigner lattice (dashed lines)
for $r_s=25$.
The dotted line is the classical plasmon dispersion. 
}
\label{fig3}
\end{figure}

In Fig. 3 we compare the plasmon dispersion calculated within the
quantum HA
with the ``phonon'' modes of the electron solid (hexagonal Wigner 
crystal (WC) \cite {thirteen}) within the harmonic approximation
upto the zone boundary of the lattice
for $r_s=25$ (corresponding to an electron density,
$n=5.4\times 10^{8} cm^{-2}$, for the 2D GaAs system). 
Inset shows the hexagonal Wigner lattice in the momentum space. 
$\omega_L$ ($\omega_T$) indicates the longitudinal (transverse) phonon mode
of the 2D Wigner lattice. The longitudinal WC phonon mode
(which corresponds to the plasmon in the electron liquid system)
has much lower frequency than the quantum
plasmon mode at $r_s=25$, which is consistent with the fact that
the transition to a WC phase is expected to occur at substantially
lower electron densities (around $r_s =37$ or below). In the long wave 
length limit the WC phonon mode
agrees with plasmon mode within HA. But in the high wave vector region
$q > k_F$  ($k_F=5.7\times 10^4 cm^{-1}$), we find that the HA plasmon
mode has a much higher frequency than the optical phonon mode.  
The experimentally measured plasmon dispersion in ref. \onlinecite {eleven}
is also much higher than the WC results shown in Fig. 3
indicating that the WC physics is unlikely to be playing 
a role here.

In considering the transition to the low density WC phase 
(and in particular whether the plasmon, or more appropriately the 
longitudinal optical phonon, mode in the electron crystal phase could 
be observed in low density 2D GaAs based electron systems) we show in 
Fig. 4 our best (approximate) estimate for the 2D electron liquid/crystal 
phase diagram (appropriate for 2D GaAs based electron systems)
in the density-temperature ($n,T$) plot. The classical 
WC - electron liquid (first order) phase transition line is defined by 
$\Gamma = \Gamma_c$, where 
$\Gamma =\langle V \rangle/\langle T \rangle$, where
$\langle V \rangle = e^2 (\pi n)^{1/2}$ is the classical
mean potential energy and 
$\langle T \rangle$ is the mean classical kinetic energy given by

\begin{figure}
\epsfysize=2.6in
\centerline{\epsffile{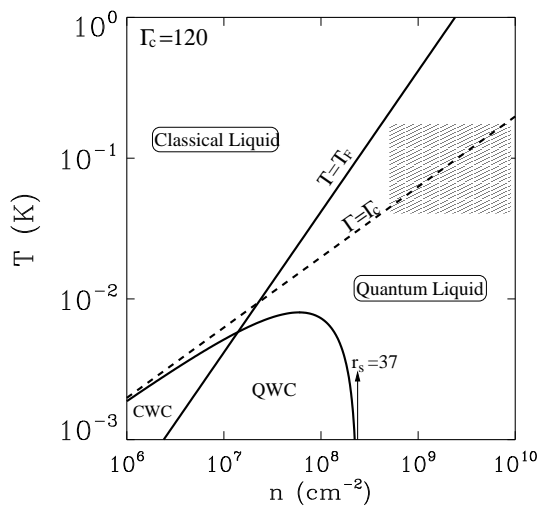} }
\vspace{0.5cm}
\caption{
Calculated 2D electron liquid - Wigner crystal 
phase diagram in the density - temperature ($n,T$) plot (
appropriate for GaAs systems).
The QWC (CWC) represents the quantum (classical) Wigner crystal.
}
\end{figure}

\[\langle T \rangle=\frac{2}{n}\int \frac{d^2p}{(2\pi^2)} 
\frac{\hbar^2p^2}{2m}n_F(p),\]
where $n_F(p)$ is the finite temperature Fermi distribution function,
and $\Gamma_c$ is found from numerical (molecular dynamics)
simulations \cite{r14} to be 
$\Gamma_c \approx 120$. The high temperature region above the $\Gamma
=\Gamma_c$ line (dashed line), $\Gamma <\Gamma_c$, 
in Fig. 4 is a classical Coulomb liquid (for $T>T_F$)
whereas the low $T$ region below the 
$\Gamma = \Gamma_c$ line, $\Gamma > \Gamma_c$, 
is the classical Wigner crystal (CWC), for $T>T_F$.
At zero (low) temperature quantum fluctuations produce a 
(first order) quantum 
phase transition between the (low density) electron crystal phase 
and the (high density) electron liquid phase. At $T=0$ this transition 
occurs at $r_s = r_s^*$ where the critical $r_s^*$ is found \cite{r15} 
by quantum Monte Carlo simulations to be around $r_s^*=37$. The high 
density region ($r_s < r_s^*$) in Fig. 4 to the right of the 
$r_s=r_s^*$ (at $T=0$) line is the quantum liquid whereas the low 
density region ($r_s>r_s^*$) along the $T=0$ line is the quantum 
Wigner crystal (QWC). We have produced the ($n,T$) phase diagram in Fig. 4 
using a simple mean field theory and Pad\'{e} approximation, which 
smoothly interpolates between the classical result $\Gamma = \Gamma_c$ 
as $n\rightarrow 0$ and the quantum result $r_s=r_s^*$ as 
$T \rightarrow 0$. The experimental regime explored in 
ref. \onlinecite{eleven} is shown as the shaded block in Fig. 4, 
and we have added the $T=T_F$ line to crudely separate the classical 
and quantum regimes. From Fig. 4 it is clear that the samples used 
in ref. \onlinecite{eleven} are essentially in the quantum electron 
liquid region --- increasing $T$ (both decreasing $n$ and 
$T$) one should be able 
to explore the classical liquid (Wigner crystal) region.
We propose that plasmon experiments be carried out in the 
lowest density samples at higher temperatures ($T\ge 4K$) in order 
to test the quantum-classical crossover in the 2D plasmon properties.

Finally in Fig. 5 we show our predicted plasmon 

\begin{figure}
\epsfysize=2.6in
\centerline{\epsffile{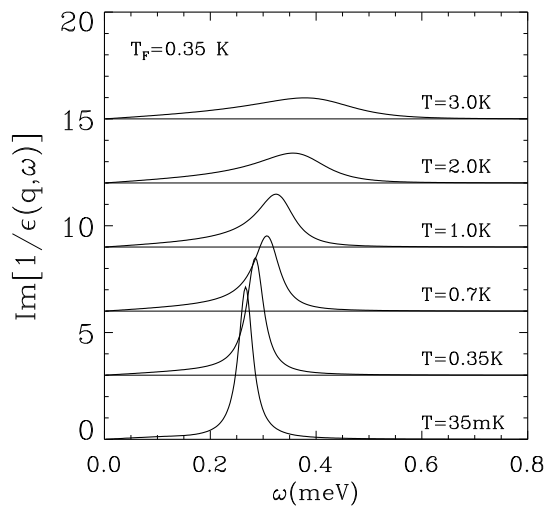} }
\vspace{0.5cm}
\caption{
The plasmon spectral weight (or the loss function, 
Im[$1/\epsilon(q,\omega)$]) at a low 2D electron density,
$r_s=18.4$ ($n=0.85 \times 10^{9} cm^{-2}$) for
various temperatures at 
a given wave vector $q=10^5cm^{-1}$. The Fermi temperature of this system
is $T_F=350 mK$. We include an impurity level broadening of $0.05meV$ in 
this calculation.
}
\end{figure}

\noindent
behavior in the classical
($T>T_F$) regime, which should be easily achievable (see Fig. 4) in the 
lowest density samples of ref. \onlinecite{eleven} simply by raising 
the temperature to 
$3-4 K$. One should observe spectacular thermal level 
broadening of the plasmon peak (as well as a temperature induced high 
energy blue shift of the plasmon energy) as shown in Fig. 5. This 
spectacular level broadening will indicate the quantum-classical 
plasmon crossover --- the current data in ref. \onlinecite{eleven} 
are all essentially still in the quantum regime. Observation of 
the WC collective modes will, however, require samples with 
substantially lower carrier densities (and lower temperature
measurements) as indicated in our phase 
diagram in Fig. 4.

We conclude by summarizing our results: We have obtained reasonable 
agreement with the measured plasmon dispersion \cite{eleven} in low 
density 2D systems by carrying out a complete quantitative calculation 
of 2D plasma dispersion. We find considerable cancellations among 
various physical mechanisms (e.g., between non-local effects and 
finite width corrections and between thermal effects and local field 
corrections) in the plasmon dispersion leading to the observed 
apparent agreement \cite{eleven} between experiment and classical 
2D plasma dispersion formula.
We also study the quantum-classical as well as crystal-liquid crossovers 
in the 2D plasmon behavior and propose specific experimental studies to
test our theoretical predictions.

This work is supported by  the U.S.-ONR.


\begin{thebibliography}{99}

\bibitem{one}S. Tomonaga, Prog. Theor. Phys. (Kyoto) {\bf 5}, 544 (1950).

\bibitem{two} D. Bohm and D. Pines, Phys. Rev. {\bf 82}, 625 91951); {\bf 85},
332 (1953); {\bf 92}, 609 (1955).

\bibitem{three}J. Lindhard, Kgl. Dan. Vidensk. Selsk. Mat. Fys. Medd. 
{\bf 28}, 8 (1954).

\bibitem{four}D. Pines and P. Nozi\`{e}res, {\it The Theory of Quantum Liquids}
(Addison-wesley, New York, 1989); G. D. Mahan, {\it Many-Particle Physics}
(Plenum, New York, 1990); N. H. March, {\it Electron Correlations in 
Molecules and Condensed Phases} (Penum, New York, 1996).

\bibitem{five} T. Ando, A. B. Fowler,
and F. Stern, \rmp {\bf 54}, 437 (1982); references therein.

\bibitem{six} D. E. Beck and P. Kumar, \prb {\bf 13}, 2859 (1976); {\bf 14},
5127 (1976); A. K. Rajagopal, \prb {\bf 15}, 4264 (1977).

\bibitem{seven} M. Jonson, J. Phys. C {\bf 9}, 3055 (1976);
N. Iwamoto, \prb {\bf 43}, 2174 (1991).

\bibitem{eight} P. M. Platzman and N. Tzoar, \prb {\bf 13}, 3197 (1976);
N. Studart and O. Hip\'{o}lito, \prb {\bf 22}, 2860 (1980).

\bibitem{nine} A. Czachor, A. Holas, S. R. Sharma, and K. S. Singwi,
\prb {\bf 25}, 2144 (1982);
A. Holas and K. S. Singwi, \prb {\bf 40}, 158 (1989).


\bibitem{ten} D. Neilson {\it et al.}, \prl {\bf 71}, 4035 (1993);
\prb {\bf 44}, 6291 (1991); S. Das Sarma and E. H. Hwang, \prl {\bf 81}, 4216
(1998).

\bibitem{eleven} M. A. Eriksson {\it et al.}, Physica E {\bf 6}, 
165 (2000); A. Pinczuk {\it et al.}, unpublished.

\bibitem{twelve} S. J. Allen, Jr., D. C. Tsui, and R. A. Logan, \prl
{\bf 38}, 980 (1977); T. N. Theis, J. P.  Kottaus, and P. J. Stiles,
Solid State Commun. {\bf 24}, 273 (1977);  T. N. Theis, Surf. Sci. 
{\bf 98}, 515 (1980).

\bibitem{r12} L. Liu {\it et al.}, Physica B {\bf 249-251},
937 (1998).

\bibitem{thirteen} R. S. Crandall, \pra {\bf 8}, 2136 (1973);
L. Bonsall and A. A. Maradudin, \prb {\bf 15}, 1959 (1977).

\bibitem{r14} R. C. Gann, S. Chakravarty, and G. V. Chester,
\prb {\bf 20}, 326 (1979);
M. Imada and M. Takahashi, J. Phys. Soc. Jpn. {\bf 53}, 3770 (1984).

\bibitem{r15} B. Tanatar and D. M. Ceperley, Phys. Rev. {\bf 39},
5005 (1989).
\end{thebibliography}
\end{document}